\documentclass[aps,pra,twocolumn,notitlepage,superscriptaddress]{revtex4-1}
\usepackage{physics}
\usepackage{xcolor,graphicx}
\usepackage{comment}
\usepackage{siunitx}

\begin{document}

\title{Experimental Quantum Embedding for Machine Learning}

\author{Ilaria Gianani}
\affiliation{Dipartimento di Scienze, Universit\`{a} degli Studi Roma Tre,
00146 Rome, Italy}

\author{Ivana Mastroserio}
\affiliation{LENS \& Dipartimento di Fisica e Astronomia, Universit\`{a} di Firenze, I-50019 Sesto Fiorentino, Italy}
\affiliation{Dipartimento di Fisica Ettore Pancini, Universit\`{a} degli Studi di Napoli Federico II, Napoli, Italy}
\affiliation{Istituto Nazionale di Ottica (CNR-INO), Largo Enrico Fermi 6, 50125 Florence, Italy}

\author{Lorenzo Buffoni}
\affiliation{LENS \& Dipartimento di Fisica e Astronomia, Universit\`{a} di Firenze, I-50019 Sesto Fiorentino, Italy}

\author{Natalia Bruno}
\affiliation{Istituto Nazionale di Ottica (CNR-INO), Largo Enrico Fermi 6, 50125 Florence, Italy}
\affiliation{LENS \& Dipartimento di Fisica e Astronomia, Universit\`{a} di Firenze, I-50019 Sesto Fiorentino, Italy}

\author{Ludovica Donati}
\affiliation{LENS \& Dipartimento di Fisica e Astronomia, Universit\`{a} di Firenze, I-50019 Sesto Fiorentino, Italy}

\author{Valeria Cimini}
\affiliation{Dipartimento di Scienze, Universit\`{a} degli Studi Roma Tre,
00146 Rome, Italy}

\author{Marco Barbieri}
\affiliation{Dipartimento di Scienze, Universit\`{a} degli Studi Roma Tre,
00146 Rome, Italy}
\affiliation{Istituto Nazionale di Ottica (CNR-INO), Largo Enrico Fermi 6, 50125 Florence, Italy}

\author{Francesco S. Cataliotti}
\affiliation{Istituto Nazionale di Ottica (CNR-INO), Largo Enrico Fermi 6, 50125 Florence, Italy}
\affiliation{LENS \& Dipartimento di Fisica e Astronomia, Universit\`{a} di Firenze, I-50019 Sesto Fiorentino, Italy}

\author{Filippo Caruso}
\affiliation{LENS \& Dipartimento di Fisica e Astronomia, Universit\`{a} di Firenze, I-50019 Sesto Fiorentino, Italy}

\begin{abstract}
The classification of big data usually requires a mapping onto new data clusters which can then be processed by machine learning algorithms by means of more efficient and feasible linear separators.
Recently, Ref. \cite{lloyd20} has advanced the proposal to embed classical data into quantum ones: these live in the more complex Hilbert space where they can get split into linearly separable clusters. Here, we implement these ideas by engineering two different experimental platforms, based on quantum optics and ultra-cold atoms respectively, where we adapt and numerically optimize the quantum embedding protocol by deep learning methods, and test it for some trial classical data. We perform also a similar analysis on the Rigetti superconducting quantum computer. Therefore, we find that the quantum embedding approach successfully works also at the experimental level and, in particular, we show how different platforms could work in a complementary fashion to achieve this task. These studies might pave the way for future investigations on quantum machine learning techniques especially based on hybrid quantum technologies.
\end{abstract}

\maketitle

\section*{Introduction}

Since ancient times transferring acquired knowledge has induced mankind to develop solutions by applying established results in different contexts. This tendency remains valid even for today's most sophisticated technologies. Indeed, the current need of processing large amount of data and the availability of supercomputers has thus fostered an innovative take on programming: rather than trying and structuring the database so that a computer can walk through it, we  now mimic processes of natural intelligence~\cite{perceptron,schmidhuber2015deep,Hastie2009}. Algorithms are then able to act as quasi-conscious agents, thanks to the introduction of machine learning. These days, its applications are ubiquitous in everyday life, ranging from domotic systems, to autonomous cars, to face and voice recognition, and to medical diagnostics~\cite{grigorescu2020survey,sebe2005machine,graves2013speech}.
\begin{figure}[t!]
\centerline{\includegraphics[width=\columnwidth]{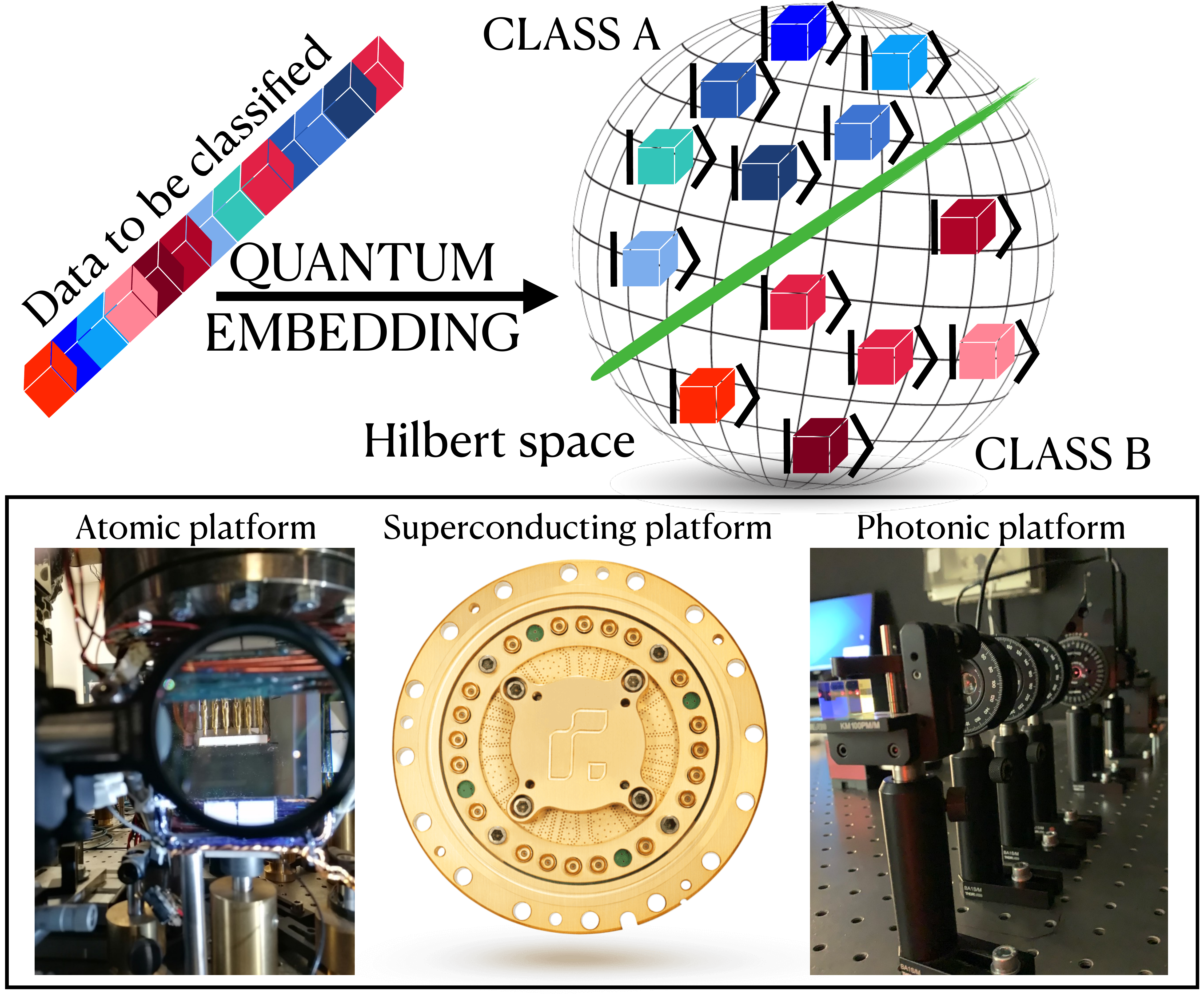}}
\caption{
Pictorial view of the quantum embedding process where classical data (originally living in a high-complex set) can be embedded into the larger (Hilbert) space of quantum states that possibly belong to tight, more distant and linearly separable clusters. We here perform several experimental tests (based on atomic, superconducting, and photonic platforms, respectively) of this quantum embedding, theoretically proposed in Ref. \cite{lloyd20}, in order to successfully demonstrate it at the experimental level and also to exploit the complementarity of the exploited platforms that is really crucial towards practical and more feasible hybrid quantum technologies. }\label{fig1}
\end{figure}
Artificial neural networks are one of the most common solutions~\cite{bishop2006pattern,goodfellow2016deep} whose exploitation targets, among the others, deep learning - the classification of data according to processes akin to those occurring in the brain - and reinforcement learning - finding the optimal strategy for a given task in complex environments~\cite{sutton2018reinforcement}. 

In classification problems, it is highly desirable that data can be sorted into distinct categories with tight dividing borders~\cite{bishop2006pattern,Hastie2009}. For this purpose, however, intensive preprocessing is often necessary on the original data for machine learning algorithms to perform efficiently. In the case of two-class problems, one would aim at achieving a geometrical representation of the data in order to establish a dividing hyperplane. All points sitting on either side of such a plane would be assigned to one class, {\it e.g.} cats or dogs, implementing what is known as a linear classifier. However, it is not granted that the natural structure of the data suits these needs. Boundary may be obvious in the geometric representation, {\it e.g.} the human-perceived distinction between cats and dogs, but, when these are applied to the original data, a complicated, non-linear classifier results. For instance, scalar data are represented as points on a line, but the classification system may require to group disjoint intervals.

Linear classifiers are a preferable option as they are easier to find, however, they require to address the nontrivial task of embedding the original data in the appropriate space. Such a preprocessing, in turn, should not be too complex and resource-intensive, with the risk of thwarting the benefits of an efficient classification. Quantum mechanics can provide an intriguing solution: even in the simplest instance, the natural representation of a quantum bit is the Bloch sphere, rather than the single-dimensional geometry of classical data. Data are thus naturally embedded into a much larger Hilbert space, and tight, separate clusters are formed, which can then be easily (quantum) recognised by a linear classifier. 

A proposal on utilising quantum computers for embedding has recently been advanced in Ref~\cite{lloyd20}. This comes in the framework of an exchange of concepts and methods between machine learning and quantum information \cite{Wittek}. Indeed, there have been demonstrations of the benefits of machine learning approaches to analyse data generated by quantum experiments~\cite{Gao18,Rocchetto19,Torlai19,Palmieri20,Giordani20,Tiunov20,Cimini20,Gebhart21}, to improve the performance of quantum sensors~\cite{Paesani17,Cimini19}, to Bayesian parameter estimations \cite{Nolan20}, to the classification of non-Markovian noise \cite{Martina21}, and to the design of optical experiments~\cite{Krenn16,Melnikov18}. Small gate-model devices and quantum annealers have been used to perform quantum heuristic optimization \cite{kadowaki1998quantum,brooke2001tunable,farhi2014quantum,peruzzo2014variational,kandala2017hardware} and to solve classification problems \cite{neven2008training,denchev2012robust,pudenz2013quantum,mott2017solving}. Such devices have even been used with good promises in the context of unsupervised and reinforcement learning \cite{otterbach2017unsupervised, vinci2020path, buffoni2021new}. These recent applications on Noisy Intermediate Size Quantum (NISQ) \cite{preskill2018nisq} devices have depicted machine learning as a good candidate to harness the power of existing quantum technologies, albeit noisy and imperfect.

The exact detail on the performance of quantum embedding will depend on the degree of control on the actual system, thus on the level of experimental imperfections specific to the solutions adopted. This does affect, foremost, the size of the classical data set which can realistically be manipulated. In this article, we present an extensive experimental study of quantum embedding carried out on multiple platforms. In particular we investigate how the protocol can be tailored to ultra-cold atoms, photonics, and via-cloud available NISQ computers. Starting from a single prescription, we implement three different experiments, highlighting requirements and tolerances of each one for this task. The specific features come into play in a complementary fashion, hence supporting the promising idea of hybrid quantum technologies for future quantum machine learning applications.

\section*{Results}

In our investigation we explore the application of quantum embedding in the simple, but illustrative instance of a single-qubit embedder. This is carried out in two steps: first, we identify the optimal quantum circuit, based on the classical data to be classified, using an iterative routine. Here the optimisation is carried out off-line on a classical computer. Secondly, the circuit is implemented in three different architectures to explore how different sources of noise and imperfections impact the realisation.

Quantum embedding is the representation of classical points $x$ from a data domain $X$ as a quantum feature state $\ket{x}$. Either the full embedding, or part of it, can be facilitated by a \textit{quantum feature map}, i.e. a quantum circuit $\Phi (x)$ that depends on the input. If the circuit has additional parameters $\theta$ that are adaptable, $\Phi (x)=\Phi (x,\theta)$, the quantum feature map can be trained via optimization.
If we have classical data points $a_i$ from class $A$, and $b_j$ from class $B$, we want their quantum embeddings $\ket{a_i}$ and $\ket{b_j}$ to be as separated as possible in the Hilbert space. The process is pictorially represented in Fig.~\ref{fig1}. The approach is similar in spirit to the classical Support Vector Machines (SVMs) commonly used in machine learning to perform classification \cite{suykens1999least}. SVMs map complex data (i.e. non linearly separable) via a nonlinear kernel into an high dimensional space where the data can be easily (linearly) classified by an hyperplane.

From the mathematical side, the determination of the optimal parameters $\theta$ of the embedding is based on computing the overlaps $\abs{\bra{a_i}\ket{b_j}}^2$, $\abs{\bra{a_i}\ket{a_{i'}}}^2$, $\abs{\bra{b_j}\ket{b_{j'}}}^2$ for all members of the two classes, and minimizing the so-called cost function C:
\begin{equation}
    C = 1 - \frac{1}{2}\left( \sum_{i,i'}\abs{\bra{a_i}\ket{a_{i'}}}^2
    +\sum_{j,j'}\abs{\bra{b_j}\ket{b_{j'}}}^2\right) + \sum_{i,j} \abs{\bra{a_i}\ket{b_j}}^2.
    \label{eq:cost}
\end{equation}
This amounts to both maximizing the Hilbert-Schmidt norm between the two classes and minimizing it within each one. This optimization procedure starts from random guesses and then leads to the optimal embeddings $\ket{a_i}$, $\ket{b_j}$, hence the associated optimal quantum circuit is found. 
In our example, the classical data set is a collection of 10 scalars $\phi$ chosen in the interval $[-\pi,\pi]$, arranged in two classes with 5 elements each (see SI for details). The optimal training is achieved by using the open-source software Pennylane \cite{pennylane} and the quantum circuit in Ref. \cite{lloyd20}, based on a sequence of rotations on non-commuting axes. The qubit is first initialised to an input-independent state $SH\ket{0}=\left(\ket{0}{+i}\ket{1}\right)/\sqrt{2}$, where $H$ is the Hadamard gate, and $S$ is the phase gate. The following sequence of rotations around the $X$ and $Z$ axes of the Bloch sphere is then applied
\begin{equation}
    \lbrace R_X(\phi),R_Z(\theta_1),R_X(\phi),R_Z(\theta_2),R_X(\phi),R_Z(\theta_3), R_X(\phi) \rbrace, 
    \label{eq:rotations}
\end{equation}
in order to construct the quantum state $\ket{\phi}$. The circuit parameters to be optimised are thus the rotation angles $\theta=\{\theta_1,\theta_2,\theta_3\}$. 

\begin{figure*}
\includegraphics[width=2.08\columnwidth]{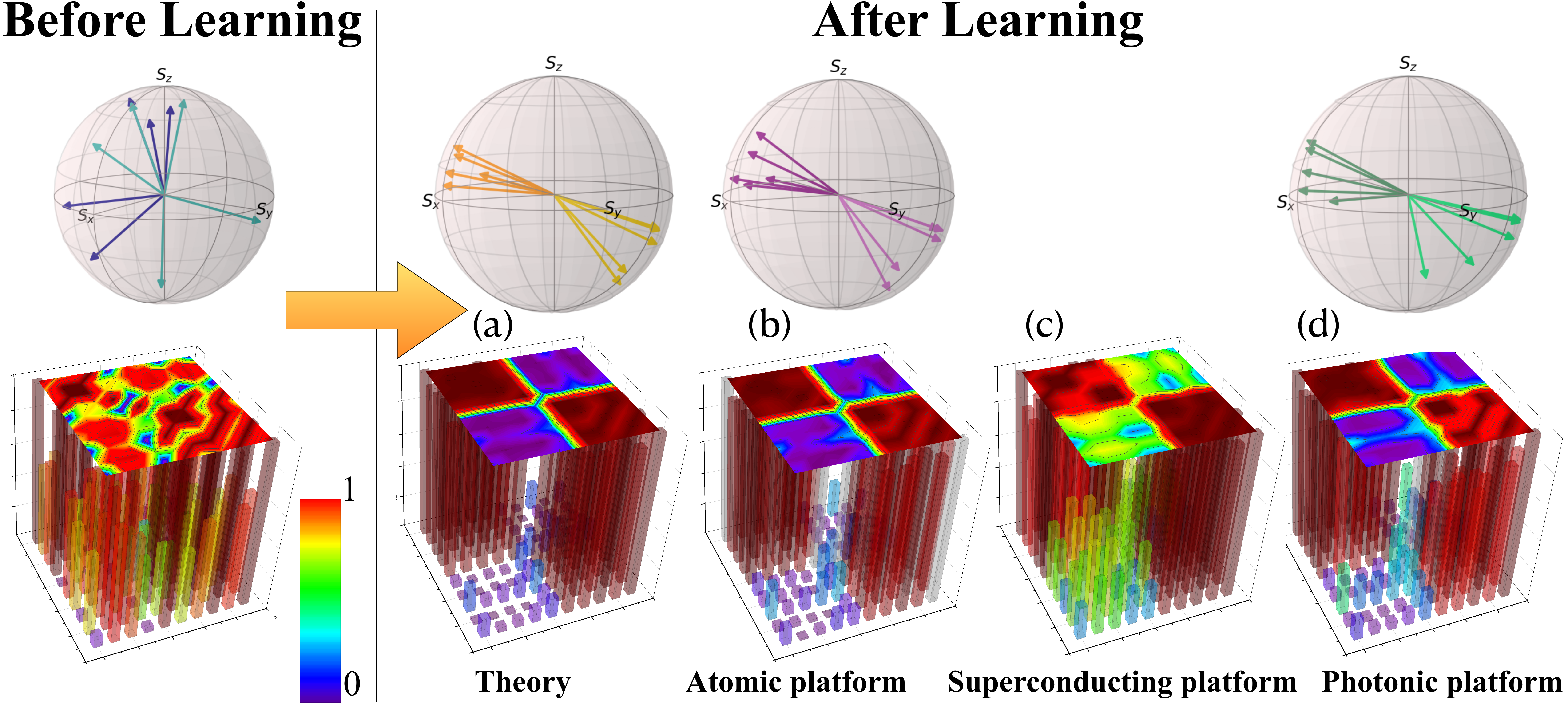}
\caption{Top panel: Bloch sphere representation of the embedded quantum states that are randomly distributed (before learning) and clustered in two families (after learning), as provided by theory and by the experimental results obtained on the atomic and photonic platforms. The corresponding Gram matrices for the set of 10 embedded quantum states (before and after learning) are shown in the bottom panel. Here, the results achieved with a superconducting platform are also shown.}\label{fig2}
\end{figure*}

The performance of the embedding is conveniently captured by the Gram matrix, containing all the scalar products between the embedded states. Fig.~\ref{fig2} shows how optimisation of the parameters $\theta$ does lead to the presence of two clusters of quantum states, as clearly illustrated by the Bloch sphere representation. As the system is trained, we can take advantage of the Gram matrix to control how well it has learned to separate the points in the Hilbert space. The training is done taking two data points, embedding them into two separate qubits and using a third qubit to perform a SWAP test between the two embeddings, hence giving us the overlaps we need to compute the cost function in Eq. (\ref{eq:cost}). The parameters of the embedding circuit are then updated by gradient descent using the automatic differentiation capabilities of Pennylane. After a few hundreds of gradient descent steps (i.e. training steps), the cost function reaches a minimum and the optimal encoding circuit is found~\cite{lloyd20} -- as illustrated in Fig.\ref{fig2}(a) for the optimally embedded (theoretical) states. 
This procedure is completely general, and allows extension to a more complex data set (see SI) and to rotations around different axes. Therefore, the scheme is flexible and can be manipulated in order to account for the specific needs of different experimental platforms. 

The manipulation of atomic internal states is naturally suited to the above scheme, since qubit operations in Eq. \eqref{eq:rotations} are indeed rotations realized by sequences of control pulses. The first experiment is therefore performed on a Bose-Einstein condensate (BEC) of approximately $10^5$ $^{87}$Rb atoms. The qubit is identified as the two level system $\left\{|0\rangle\equiv|F=2, m_F=0\rangle;|1\rangle\equiv|F=1, m_F=0\rangle\right\}$ formed by rubidium hyperfine ground states (see SI Fig.\ref{SI- atomsetup}).
The classical information to be encoded is mapped in the rotation angles of the Bloch vector of our two-level system, around the non commuting axes $S_x$ and $S_z$ of the Bloch sphere. The sequence of rotations necessary to realize the embedding is thus implemented by a series of quasi-resonant microwave pulses applied to the atoms.
The Bloch vector representing our qubit will evolve, under the rotating wave approximation, according to the unitary transformation:
\begin{equation}
U_{\Omega}(\Omega,\delta,t)=exp\left(-\frac{i|\vec{\Omega}|t}{2}\vec{n}_\Omega\cdot\vec{\sigma}\right)
\end{equation}
with $|\vec{\Omega}|=\sqrt{\Omega^2+\delta^2}$ being the generalized Rabi frequency and $\vec{n}_\Omega\equiv\left(\frac{\Omega}{|\vec{\Omega}|},0,-\frac{\delta}{|\vec{\Omega}|}\right)$ identifying the rotation axis. Here, $\vec{\sigma}$ is the vector of the Pauli matrices, $\delta$ is the detuning of the microwave frequency from the atomic resonance, while $\Omega$ is the Rabi frequency of the pulse. Therefore, when the microwave pulse is off ($\Omega=0$), the system rotates around the $z$ axis with an angular frequency equal to the detuning $\delta$. Since we cannot change the detuning during the evolution, when the microwave pulse is on ($\Omega\neq0$) the system rotates with an angular frequency equal to the generalized Rabi frequency $|\vec{\Omega}|$ around an axis that forms an angle of $\arctan\left(\delta/\Omega\right)$ with the $x$ axis. A correction factor on the interaction time must then be introduced in computing the sequence of rotations due to the non-orthogonality of the two rotation axes. The sequence is therefore reduced to:
\begin{equation}
\left\{R_{x}(\varphi_{1}), R_{z}(\vartheta), R_{x}(\varphi_{2})\right\}
\end{equation}
where the angles $\varphi_{i}=\Omega\tau_{i}$ are determined by the interaction time $\tau$ when the microwave is switched on, while the time interval $T$ when the microwave is switched off determines the angle $\vartheta=\delta T$.
To reconstruct the final embedded state after the total evolution, we perform a state tomography as explained in detail in the SI.
The sources of error and imperfections in the implementation of the protocol on the atomic experimental platform come from the limited accuracy and stability of the measured Rabi frequency and detuning used to set the desired rotations.
As explained in the SI this induces an error of about 0.1 in the measurement of the components $S_x$, $S_y$ and $S_z$ of the final collective-spin vector $\vec{S}$.
The experimental findings in terms of the Bloch vectors and Gram matrix are reported in Fig.\ref{fig2}(b) where a clear clusterization of the reconstructed states can be seen. The fidelity between the theoretical prediction and the experimental result is on average better than 0.99 (see Fig.\ref{SI-atomfidelities} in the SI). We can safely conclude that the embedding procedure is faithfully working with the atomic platform.

Different platforms, however, do not necessarily have the same features and constraints. We  evidence this, by taking as an example the case of the superconducting chip of Rigetti \cite{rigettiqpu} named Aspen-8. It is composed by a lattice of 30 superconducting qubits in a ladder-like configuration. The qubits are controllable by the action of single and two-qubit gates giving this platform all the characteristics of an universal quantum computer. We thus use this platform to deploy again the same embedding in Eq.(\ref{eq:rotations}) but on a completely different device with different noise profile and constraints. Each circuit is sampled $2000$ times for each of the $100$ datapoints necessary to build the Gram matrix, using Rigetti's cloud service. As shown in Fig.\ref{fig2}(c), the Gram matrix (even if it is sensibly noisier than our other tests) clearly exhibits the separation boundary between the two classes. The advantage of this experiment is that it could be performed without the need of an ad-hoc lab; it has been carried out remotely by just reserving some time on the Rigetti system and programming it \cite{smith2016practical}. The entire set of experiments performed on this platform to get these results has taken a total time of around $5$ minutes to run.

The examples above detail results obtained on platforms with static material qubits. It is also of interest to perform the quantum embedding with moving qubits \textit{i.e.} photons.
Thus, we have also investigated quantum embedding in an optical experiment by generating photon-pairs at the degenerate wavelength \SI{810}{\nano m} via parametric down conversion realising a heralded single-photon source. The input state of the photons is set to $|H\rangle$ and then the embedding is performed by applying a single unitary:
\begin{equation}
\label{eq:Ufotonica}
    \tilde{U}(\phi;\vec n)=\cos(\phi/2)\sigma_0-i\sin(\phi/2)(\vec{n} \cdot \vec{\sigma})
\end{equation}
where $\vec{n}=(n_x,n_y,n_z)$ denotes the associated rotation axis (the actual used values are in the SI).
This alternative approach of constructing the embedding in the form of a feasible unitary $\tilde{U}(\phi;\vec{n})$, also including the $SH$ initialisation, is more suited to photonic experiments. On one hand, this scheme requires superior controllability of the system, since we cannot choose the rotation axis in advance. On the other, using only a one-shot embedding instead of a step-by-step approach is a significant simplification. 
For photon polarisation qubits, rotations in the form \eqref{eq:Ufotonica} are conveniently performed by means of a series of three wave plates - a quarter wave plate (Q1), a half wave plate (H2) and a second quarter wave plate (Q3), whose angles are
associated to the parameters $\phi$ and $\vec n$ as
\begin{eqnarray}
&\theta_{Q1}=\frac{1}{2} \left(-\arctan \left(\frac{n_z}{n_x}\right)-\arctan\left(n_y \tan \left(\frac{\phi }{2}\right)\right)\right) \nonumber \\
&\theta_{H2}=\frac{1}{2} \left(-\arcsin\left(n_x \sqrt{\frac{n_z^2}{n_x^2}+1} \sin \left(\frac{\phi }{2}\right)\right)-\arctan\left(\frac{n_z}{n_x}\right)\right) \nonumber\\
&\theta_{Q3}=\frac{1}{2} \left(-\arctan\left(\frac{n_z}{n_x}\right)-\arctan\left(n_y \tan \left(\frac{\phi }{2}\right)\right)\right) \nonumber \; .
\end{eqnarray}\\
This is a standard approach, relying on the availability of reliable hardware components, at the cost of losing a direct mapping of each term in the sequence 
\eqref{eq:rotations} into a physical object. Verification of the outputs is carried out by means of quantum state tomography, operated by a further half-wave plate/ quarter-wave plate sequence and a polariser.
The reconstructed Gram matrix is reported in Fig.~\ref{fig2}(d). The two main sources of imperfection are the limited accuracy in the setting of the wave-plate axis ($\pm 1^{\circ}$ for the encoding plates, $\pm 0.25^{\circ}$ for the tomography plates), and deviations of the imparted phase shifts from the target values $\pi/2$ or $\pi$. Despite these deviations, clustering of the states in two classes is clearly observed. Once again the fidelity between the experimental results and the theoretical predictions is on average above 0.96 (see Fig. \ref{SI-opticsfidelities} in the SI) accounting for a more than satisfactory agreement.

\section*{Discussion}

The Gram matrices reveal that all the architectures we have considered indeed give satisfactory results in realising the quantum embedding. To make this observation more quantitative, we here provide a theoretical bound, based on the observed fidelities, for the number of different classes and/or different points that can be embedded on a single qubit. 
Our bound is derived from geometrical constraints of the embedding on the sphere. The minimal assumption we can make for the embedding to be successful is to avoid overlapping between different classes. These are associated to distinct spherical sectors defined by its central angle $\theta$; in our example, the two sectors correspond to two separate halves of the sphere ($\theta=\pi$). By simple geometrical arguments, the surface occupied by a single spherical sector is given by $2\pi [1-\cos(\theta/2)]$. Since we want the sectors to be non-overlapping, and the total available surface of our Bloch sphere is $4\pi$, we can formulate a geometrical constraint considering the number of classes $N$ as follows:
\begin{equation}
    2\pi N\left[1- \cos\left(\frac{\theta}{2}\right)\right] \leq 4\pi.
    \label{eq:sphere}
\end{equation}
This equation thus computes the tightness of the embedding given a number $N$ of classes to embed. 

These methods also provide an upper bound on the data set which can effectively be embedded. We use the fidelity $F$ between the experimental states and the corresponding targets. While the ideal state is associated to a unique direction in the Bloch sphere, accounting for the imperfections of the actual prepared state leads us to consider a spherical surface $2\pi(1-F)$ as the proper geometric measure. Thus, the maximum number of points allowed is
\begin{equation}
N_{\rm max}\leq\frac{4\pi}{2\pi(1-F)}.
    \label{eq:maxpoints}
\end{equation}
Our experiments demonstrate that a fidelity exceeding $0.9$ can be routinely achieved on all platforms we have explored. This determines a maximal embedding capacity $N_{\rm max}\simeq 20$ as a conservative estimate. This reasoning can be obviously generalized to the $2^n$-dimensional hypersphere in the case of multi-qubit embedding.

Our experimental investigation demonstrates how quantum embedding techniques may suit radically different approaches to qubit encoding and manipulation - by pulses as for cold atom, by quantum logic circuits as for the Rigetti machine, or by compiled operations as for photons. Such a versatility shows promises for future interconnected systems on hybrid architectures, with specialised hardware for storage, processing and distribution of quantum data. Finally, the potential advantages of representing classical data on quantum systems include not only the possibility to simplify a classification problem as experimentally demonstrated in this work, but also the ability to speed up any processing of the classical data such as, among the others, the quantum parallelism to search through a database, feature extraction, image segmentation, and edge detection. Indeed, combining quantum machine learning and quantum image processing is expected to allow to potentially solve real-world problems that are very challenging via classical supercomputers, especially in the case of large volumes of data in various domains ranging from sociology to economy, from geography to biomedicine.

\section*{Authors contribution}
I.G., V.C. and M.B. carried out the photonic experiment; I.M., N.B., L.D. and F.S.C. carried out the atomic experiment; F.C. and L.B. led and carried out the theoretical work. L.B. performed the numerical optimizations and the analysis of the results from the Rigetti machine. F.C. conceived the whole project. F.C., M.B. and F.S.C. supervised the project. All authors contributed to the discussion, analysis of the results and the writing of the manuscript.

\section*{Acknowledgments}
F.C. acknowledges funding from Fondazione CR Firenze through the project Quantum-AI, from Florence Univ. under grant Q-CODYCES, and from European Commission through the Programme H2020 EU FET-OPEN project PATHOS grant no. 828946. N.B., L.D., I.M. and F.S.C. were financially supported from Qombs Project, FET Flagship on Quantum Technologies grant n. 820419.

\section*{Methods}

\subsection*{Atomic platform}

Starting from a room temperature gas in ultra-high vacuum conditions we produce in \SI{8}{s} a BEC of $^{87}$Rb atoms by laser cooling  followed by evaporative cooling in a magnetic micro-trap realized with an atom-chip \cite{atom_chip}. Quantum degeneracy is achieved by ramping down the frequency of a radio-frequency field integrated on the chip. This procedure yields a BEC of approximately $10^5$ atoms in $\left|F=2,m_F=2\right\rangle$ (see SI), at a critical temperature of \SI{0.5}{\micro K} and a distance of \SI{300}{\micro m} between the atomic cloud and the chip surface. After switching off the magnetic trap, we let the atoms expand for 1 ms to strongly reduce the effects of atomic collisions. Subsequently, we apply a constant magnetic bias field of 6.179 G to lift the magnetic degeneracy of the hyperfine states and to define the quantization axis of the system. The opposite sign of the Land\'{e} factors in the two hyperfine ground levels isolates the $\left|F=2\right\rangle \rightarrow \left|F=1\right\rangle$ two-level system microwave clock transition chosen as our qubit for the embedding. To transfer all the atoms in the initial state of our experiment $\left|0\right\rangle=\left|F=2,m_F=0\right\rangle$, we apply a frequency modulated radio-frequency pulse designed with an optimal control strategy \cite{opt_control}. We then exploit an external microwave-antenna at \SI{6.834}{\giga Hz} to drive the clock transition inducing an average Rabi oscillation frequency of $2 \pi \times$\SI{38(2)}{\kilo Hz}.
After all the manipulations to realize the embedding of the states, we finally record the number of atoms in each of the $m_F$ Zeeman states, of the $\left|F=1\right\rangle$ and $\left|F=2\right\rangle$ hyperfine levels, applying a Stern-Gerlach method \cite{Stern_Gerlach}. We let the different $m_F$ states to spatially separate by means of an inhomogeneous magnetic field applied along the quantization axis and then, after 23 ms of free expansion, we execute a standard absorption imaging sequence. The sequence is repeated twice, once with light resonant with the $\left|F=2\right\rangle$ state and immediately after with light resonant to the $\left|F=1\right\rangle$ state.
%
%
\subsection*{Photonic platform}

Degenerate  810  nm  photon  pairs  are  generated  from a 50 mW CW laser at 405 nm through Type I Spontaneous  parametric  downconversion  using  a  3mm  BBOcrystal.  Both photons are filtered with bandpass filters (FWHM=7.3 nm) and single mode fibers.  One photon acts as a trigger and is directly coupled to an Avalanche Photodiode  Detector, while the other undergoes the transformation $\tilde{U}(\phi;\vec{n})$ imparted by the sequence of quarter - half - quarter wave plates described in the main text.  Quantum state tomography is then performed by collecting coincidence counts in correspondence of the projections along the polarisation directions Horizontal, Vertical, Diagonal, Anti-diagonal, Right-circular and L-circular. For each state, approximately $20k$ coincidence events are collected, distributed among the six projectors.


\section*{Supplementary Information}

\subsection*{Theoretical details}

\begin{figure}[b!]
    \centering
    \includegraphics[width=0.85\linewidth]{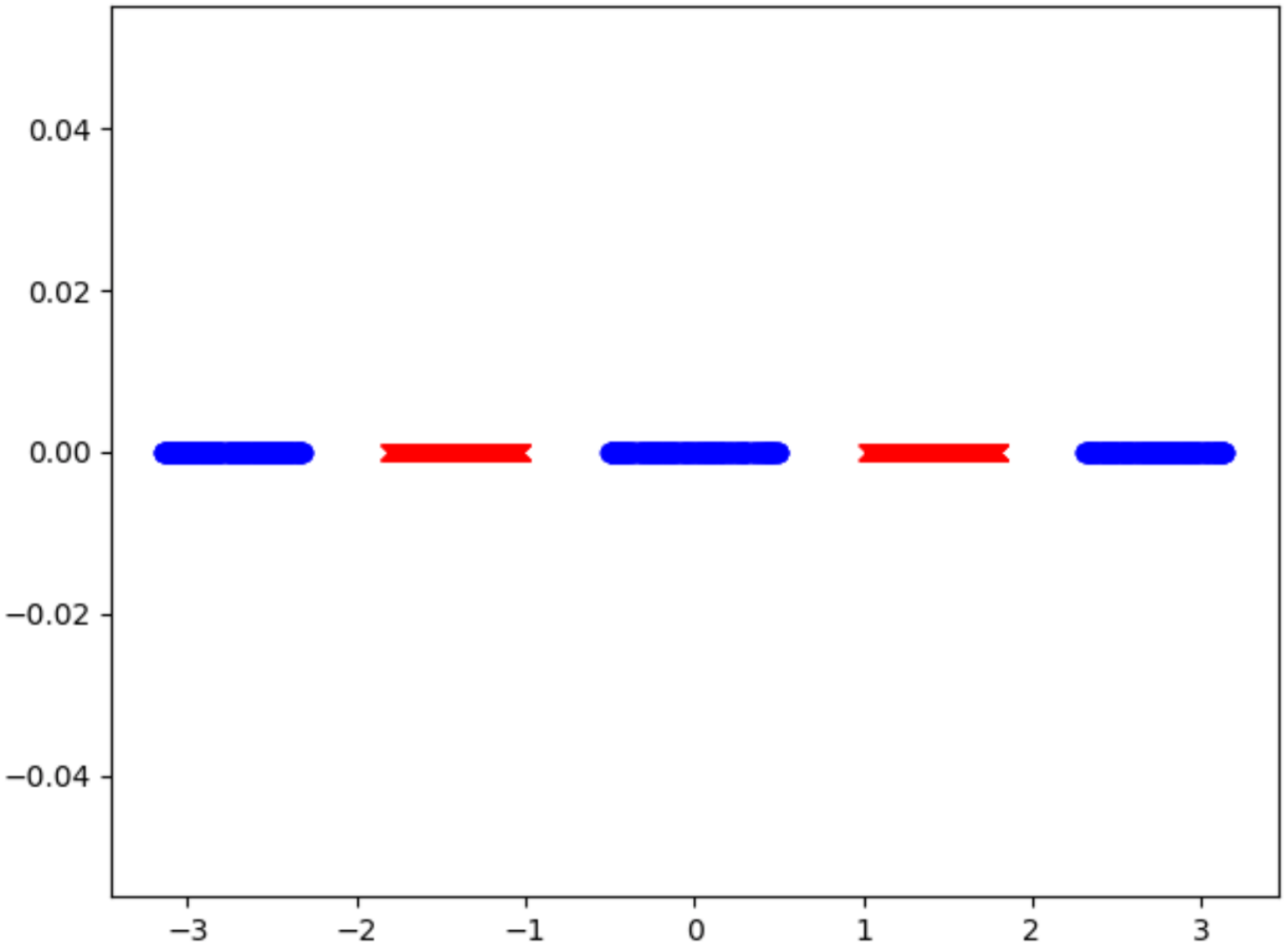}
    \caption{Data set to be classified using the embedding. The data set is one-dimensional and it is not linearly separable. The classes (blue) and (red) have been normalized to live in the interval $[-\pi, \pi]$.}
    \label{SI-Fig1}
\end{figure}

\begin{figure}[h!]
\centering
\includegraphics[width=1\linewidth]{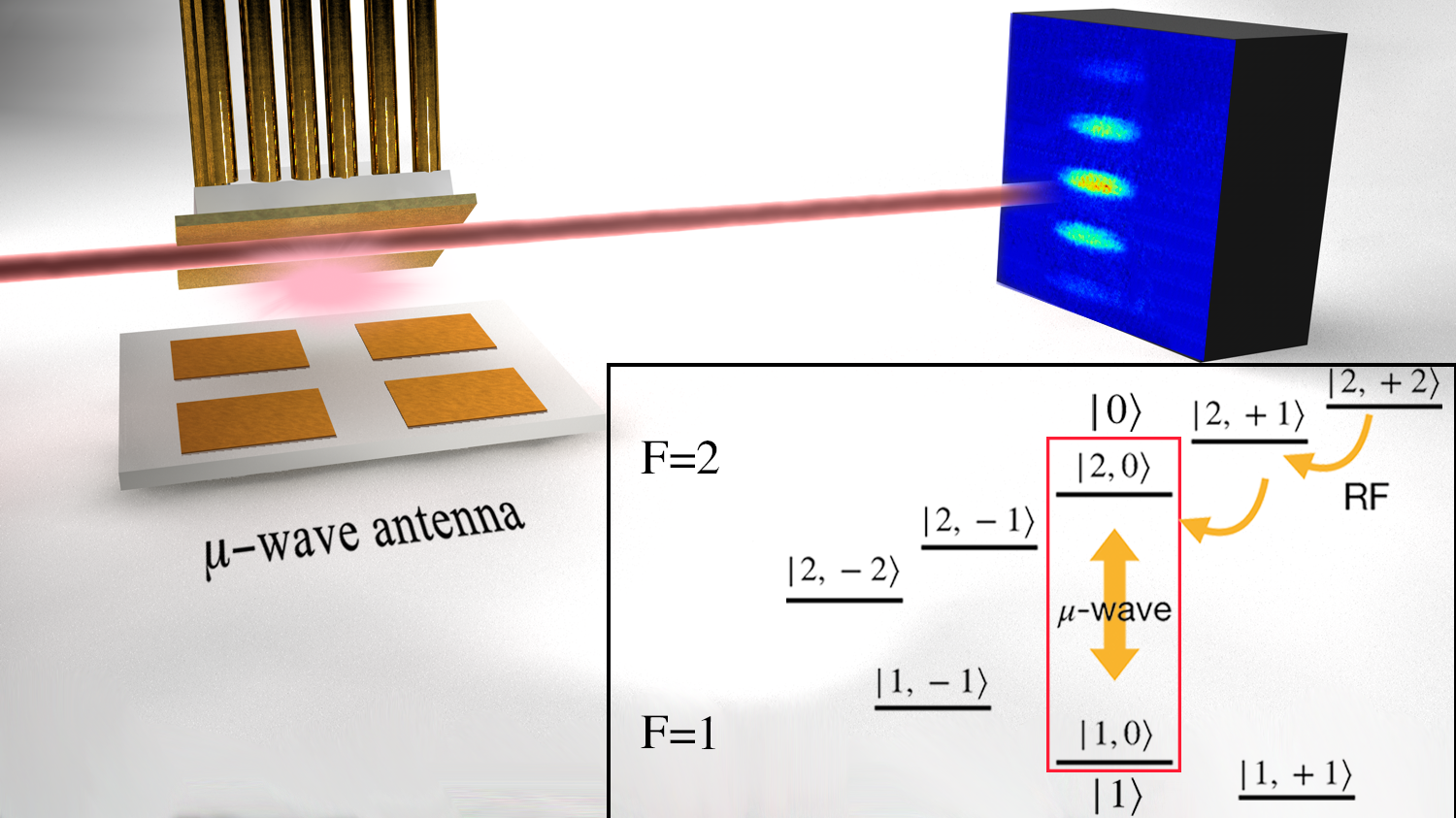}
\caption{Left: Pictorial representation of the atom-chip placed inside the vacuum science cell to achieve the BEC and manipulate the atoms internal dynamics. An external microwave antenna, shown in the image below the chip, is used to drive the evolution of the qubit and to realize the embedding. Note that the drawing is not to scale. Right: Typical absorption image of the atomic clouds after Stern-Gerlach separation. All Zeeman substates can be simultaneously detected. In the inset we show the involved level scheme with yellow thin lines representing the radio-frequency transitions while thick lines represent the microwave transition.}
\label{SI- atomsetup}
\end{figure}

The classical data set is reported in Fig. \ref{SI-Fig1}. This data set cannot be linearly separated in one-dimensional representation, thus making the quantum embedding a suitable resource to classify the two classes. In our case the training is performed on $1000$ points and the cost function is optimized as explained in the main text by using the simulation software Pennylane. The peculiarity of Pennylane is that it integrates a suite of quantum simulations (that we use to implement the circuit) with common machine learning tools such as TensorFlow and Pytorch. Leveraging the remarkable automatic differentiation and optimization capabilities of these libraries our procedure is able to converge to a minimum of our cost function in $200$ iterations taking only a few minutes of computational time. We use the TensorFlow backend with the Adam optimizer.

Once the optimization is done, we sample $10$ more data points (not exploited for the training) in order to test the generalization capabilities of the embedding. These $10$ test points are all correctly classified and they subsequently serve as a  benchmark for all the experimental platforms. After this "learning" phase has been carried out on an ideal circuit, we deploy the same embedding on different experimental platforms to test its robustness to real-world deployment scenarios. Since different experiments have different capabilities, each deployment requires to 'recompile' the embedding unitary transformation to fit with the specifications of each device while keeping the learned parameters fixed. The different constraints are extensively explained in the following sections.

\subsection*{Atomic experiment}
\textbf{Experimental procedure.}
The embedding is realized by letting the Bloch vector evolve from the initial state $\left|0\right\rangle=\left|F=2,m_F=0\right\rangle$ to the final embedded state $\left|x\right\rangle$ through the set of rotations $\left\{R_{x}(\varphi_{1}), R_{z}(\vartheta), R_{x}(\varphi_{2})\right\}$ dictated by the embedding. The latter are implemented designing a microwave rectangular wave, whose duration at minimum and maximum is fixed by the angles $\varphi_{i}=\Omega\tau_{i}$ and $\vartheta=\delta T$, where $\Omega = 2 \pi \times$\SI{38(2)}{\kilo Hz} is the Rabi frequency and $\delta = 2\pi\times$\SI{6.57(4)}{\kilo Hz} the detuning of the microwave frequency from the atomic resonance. The interaction time with microwave switched on ($\tau_1$,$\tau_2$) and with microwave switched off ($T$) are reported in Table \ref{SI:atomicsparameters}.

  \begin{table}[h!]
  \setlength{\tabcolsep}{1.2em}
  \renewcommand{\arraystretch}{1.5}
  \begin{tabular}{|  c  |  c  |  c  | c |}
    \hline
		$\,\,State\,\,$ & $\tau_1 [\mu s]$ & $T [\mu s]$ & $\tau_2 [\mu s]$ \\ \hline
		\hline
		1 & 19 &	36 &	8 \\ \hline
		2 & 45 &	51 &	7  \\ \hline
		3 & 20 &	37 &	1 \\ \hline
		4 & 20 &	38 &	3 \\ \hline
		5 & 19 &	28 &	3 \\ \hline
		6 & 30 &	47 &	38 \\ \hline
		7 & 32 &	20 &	8  \\ \hline
		8 & 4  &	7  &	35 \\ \hline
		9 & 4  &	25 &	12  \\ \hline
		10 & 7 &    10 &	6 \\ \hline
        \hline
  \end{tabular}
  \caption{Interaction times with the microwave switched on ($\tau_1$,$\tau_2$) and off ($T$) for the atomic experiment, in order to generate the $10$ optimally embedded quantum states.}
  \label{SI:atomicsparameters}
  \end{table}
  
At the end of the evolution, we measure the relative population $R_p=P_{\left|1\right\rangle}/(P_{\left|0\right\rangle} + P_{\left|1\right\rangle})$ of each final state and we get  $\left\langle S_z \right\rangle$ as
\begin{equation}
    \left\langle S_z \right\rangle = \frac{P_{\left|0\right\rangle} - P_{\left|1\right\rangle}}{P_{\left|0\right\rangle} + P_{\left|1\right\rangle}} \; ,
\label{eq:Sz}
\end{equation}
where $P_{\left|0\right\rangle}$ stands for the population of the state $\left|0\right\rangle$ and $P_{\left|1\right\rangle}$ for the state $\left|1\right\rangle$. To measure the other components $S_x$ and $S_y$ of the final state, we project them onto the z axis of the Bloch sphere by adding further rotations to the one computed to measure $S_z$. Experimentally, we realize them by adding another sequence of rectangular wave pulses of suitable duration.
In this way we retrieve $S_x$ and $S_y$ by the measured relative population with Eq.(\ref{eq:Sz}). By doing so, we implement the quantum state tomography of the embedded states.

\begin{figure}[h!]
\centerline{\includegraphics[width=\columnwidth]{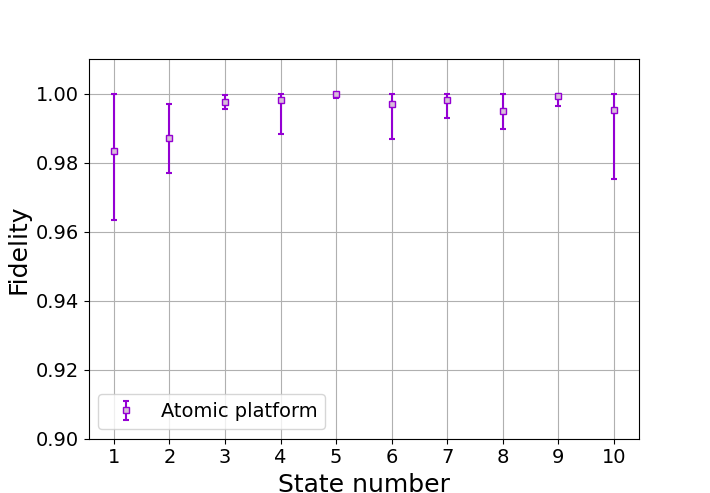}}
\caption{Fidelity between the predicted state and the one experimentally reconstructed one in the atomic experiment by measuring the three $\vec{S}$ components after applying the quantum embedding circuit, calculated for the 10 validation states. Each measurement of the components is repeated 5 times and the uncertainty on the fidelity is obtained via error propagation. The error bar for the state 5 is smaller than the marker.}\label{SI-atomfidelities}
\end{figure}

\textbf{Accuracy and stability.} The experimental realization of the embedding protocol with atoms is fundamentally limited by the accuracy and stability with which we measure the Rabi frequency and the radio-frequency detuning necessary to set the desired rotations. Therefore, in order to estimate this accuracy, we have performed Rabi oscillations and a Ramsey interferometer up to a duration of \SI{300}{\micro s}, much longer that any of the embedding sequences. We have hence measured a fluctuation of the order of 1\% on Rabi and detuning frequencies. Furthermore, the frequency stability, presumably due to microwave power fluctuations for the Rabi frequency and to magnetic field fluctuations for the detuning, have been evaluated measuring the Allan variance for a total time interval of three hours. The results have allowed us to estimate a fluctuation of $2\pi\times$ \SI{1.5}{\kilo Hz} and of $2\pi\times$ \SI{71}{Hz} for Rabi frequency and detuning, respectively, evaluated on an averaging time of 115 s corresponding to the time needed for five repeated measurement of each embedded state. The linear uncertainty propagation realized with a simulation of the experiment results in an evaluated error of 0.1 on the measurement of the components $S_x$, $S_y$ and $S_z$ of the final collective-spin vector. The fidelities between the predicted states and the reconstructed experimental ones, together with error bars computed via uncertainty propagation, are shown in Fig. \ref{SI-atomfidelities}.
%
%
%

\subsection*{Photonic experiment}

\begin{figure}[h!]
\includegraphics[width=0.9\linewidth]{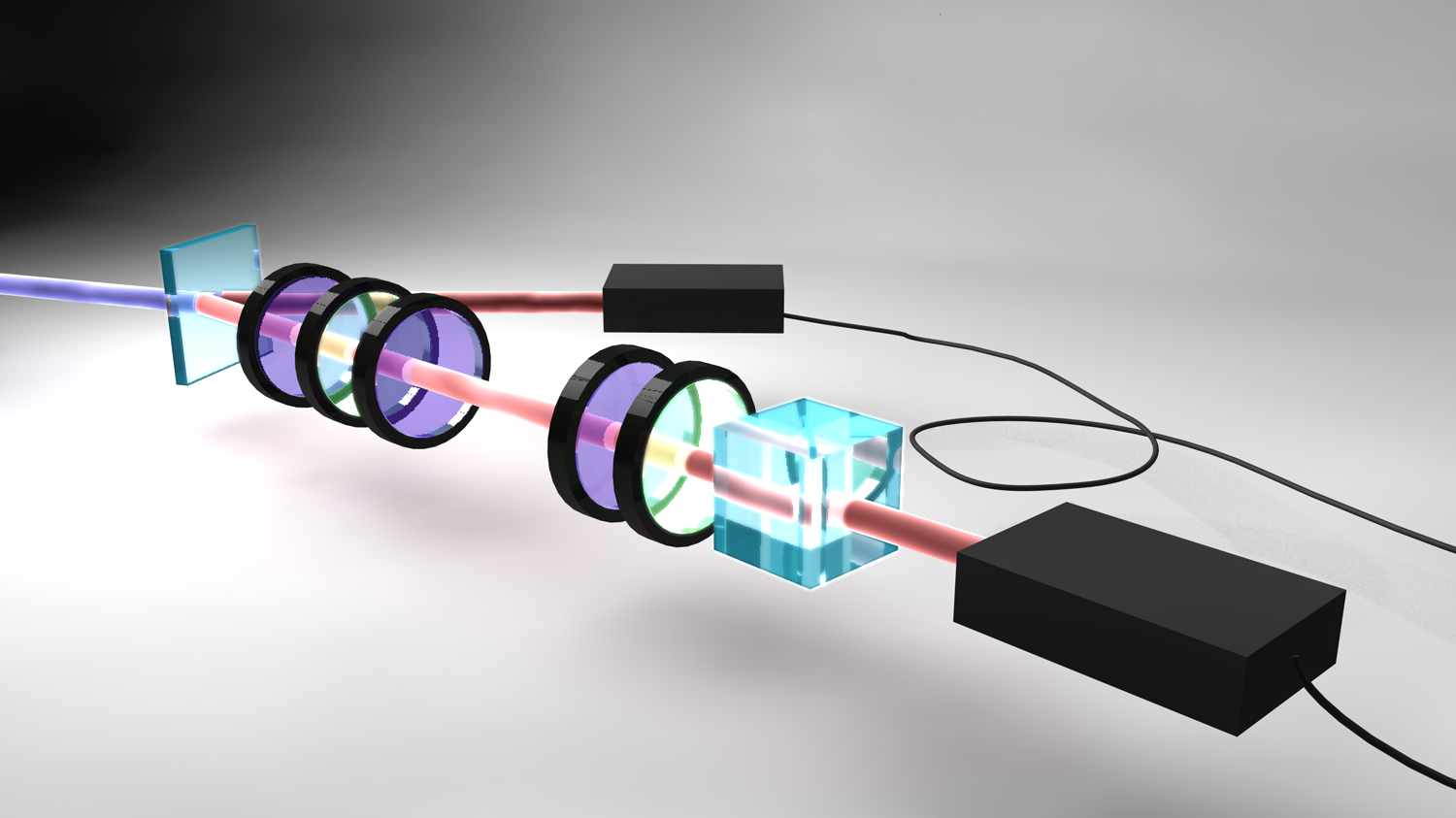}
\caption{Optical experimental setup: SPDC through a 3mm Type I BBO crystal generates degenerate photon pairs at 800 nm. One photon is used for heralding, while the other is rotated with a Quarter wave plate - Half wave plate -Quarter wave plate and then quantum state tomography is performed with a Quarter wave plate, Half wave plate and a Polarizing beam splitter.Coincidence counts are then collected as reported in the main text.}
\label{SI- opticssetup}
\end{figure}

The experimental setup used for the photonic experiment is shown in Fig. \ref{SI- opticssetup}. A heralded single photon source is set up, based on parametric down conversion, and a three-plate arrangement is used to impart a generic transform, while tomography adopts the standard technique of projecting on three mutually unbiased bases.

  \begin{table}[h!]
  \setlength{\tabcolsep}{1.2em}
  \renewcommand{\arraystretch}{1.5}
  \begin{tabular}{|  c  |  c  |  c  | c | c |}
    \hline
		$\,\,State\,\,$ & $\phi$ & $n_x$ & $n_y$ & $n_z$ \\ \hline
		\hline
		1 & 0.668 &	0.667 &	0.143 & 0.731 \\ \hline
		2 & 1.986 &	-0.423 &	0.460 & -0.781 \\ \hline
		3 & 2.111 &	-0.510 &	0.379 & -0.772 \\ \hline
		4 & 2.408 &	0.619 &	0.240 & 0.748 \\ \hline
		5 & 1.301 &	-0.405 &	0.914 & 0.034 \\ \hline
		6 & 4.258 &	0.418 &	0.908 & -0.006 \\ \hline	7 & 4.367 &	0.247 &	0.969 & 0.026 \\ \hline
		8 & 3.549 &	-0.475 &	0.847 & 0.239 \\ \hline
		9 & 4.379 &	0.197 &	0.980 & 0.036 \\ \hline
		10 & 3.762 & -0.433 &	0.877 & 0.208 \\ \hline
        \hline
  \end{tabular}
  \caption{Embedding parameters for the photonic experiment.}
  \label{SI:opticsparameters}
  \end{table}
\begin{figure}[h!]
\includegraphics[width=0.9\linewidth]{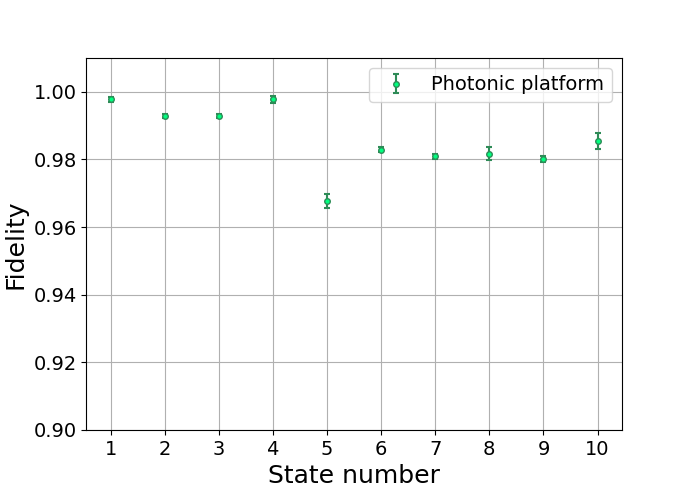}
\caption{Fidelity between the predicted state and the experimentally reconstructed one in the photonic experiment, obtained by quantum state tomography after applying the quantum embedding circuit, calculated for the 10 validation states.}\label{SI-opticsfidelities}
\end{figure}

The embedding parameters dictating the rotation for each of the 10 states are reported in Table \ref{SI:opticsparameters}. 
The fidelities between the predicted states and the reconstructed ones are shown in Fig. \ref{SI-opticsfidelities}. Uncertainties have mainly statistical origin, due to fluctuations in the number of collected counts. In order to account for those, we have performed a Monte Carlo routine, simulating 300 experiments: in each of them, counts are generated by bootstrapping on the observed values, assuming Poissonian statistics, as customary. The uncertainty on the fidelity is thus evaluated as the standard deviation of the fidelities with the target state observed in the Monte Carlo replicas.




\begin{thebibliography}{99}

\bibitem{perceptron}
F. Rosenblatt, The perceptron: A probabilistic model for information storage and organization in the brain. Psychol. Rev. 65 (1958).

\bibitem{schmidhuber2015deep}
J. Schmidhuber, Deep learning in neural networks: An overview. Neural Networks 61 (2015).

\bibitem{Hastie2009}
T. Hastie, R. Tibshirani, and J. Friedman, The elements of statistical learning: data mining, inference, and prediction. Springer Science \& Business Media (2009).

\bibitem{grigorescu2020survey}
S. Grigorescu, B. Trasnea, T. Cocias and G. Macesanu, A survey of deep learning techniques for autonomous driving. Journal of Field Robotics 37 (2020)

\bibitem{sebe2005machine}
N. Sebe, I. Cohen, A. Garg and T. S. Huang, Machine learning in computer vision. Springer Science \& Business Media (2005)

\bibitem{graves2013speech}
A. Graves, A. Mohamed and G. Hinton, Speech recognition with deep recurrent neural networks. ICASSP (2013)

\bibitem{goodfellow2016deep}
I. Goodfellow, Y. Bengio, A. Courville, and Y. Bengio, Deep Learning. MIT press Cambridge (2016).

\bibitem{bishop2006pattern}
C. M. Bishop, Pattern recognition and machine learning. Springer(2006).

\bibitem{sutton2018reinforcement}
R. S. Sutton and A.G. Barto, Reinforcement learning: An introduction. MIT press (2018).

\bibitem{lloyd20}
S. Lloyd, M. Schud, A. Ijaz, J. Izaac, and N. Killoran, Eprint arXiv:2001.03622 (2020).

\bibitem{Wittek}
P. Wittek, Quantum machine learning: what quantum computing means to data mining (Academic Press, 2014).

\bibitem{Gao18} 
J Gao, et al., Experimental Machine Learning of Quantum States. Phys. Rev. Lett. 120 , 240501 (2018).

\bibitem{Rocchetto19} 
A. Rocchetto, S. Aaronson, S. Severini, G. Carvacho, D. Poderini, I. Agresti, M. Bentivegna, and F. Sciarrino, Experimental learning of quantum states, Sci. Adv. 5, eaau1946 (2019).

\bibitem{Torlai19} 
G. Torlai, et al.  Integrating Neural Networks with a Quantum Simulator for State Reconstruction, Phys. Rev. Lett. 123, 230504 (2019).

\bibitem{Palmieri20}
A. M. Palmieri, E. Kovlakov, F. Bianchi, D. Yudin, S. Straupe, J. D. Biamonte, and S. Kulik, Experimental neural network enhanced quantum tomography, Npj Quantum Inf. 6, 20 (2020).

\bibitem{Giordani20} T. Giordani, A. Suprano, E. Polino, F. Acanfora, L. Innocenti, A. Ferraro, M. Paternostro, N. Spagnolo, and F. Sciarrino, Machine Learning-Based Classification of Vector Vortex Beams, Phys. Rev. Lett. 124, 160401 (2020).

\bibitem{Tiunov20} E. S. Tiunov, V. V. Tiunova (Vyborova), A. E. Ulanov, A. I. Lvovsky, and A. K. Fedorov, Experimental quantum homodyne tomography via machine learning, Optica 7, 448 (2020).

\bibitem{Cimini20} V. Cimini et al., Neural networks for detecting multimode Wigner negativity, Phys. Rev. Lett. 125, 160504 (2020).

\bibitem{Gebhart21} V. Gebhart, et al., Identifying nonclassicality from experimental data using artificial neural networks, Eprint arXiv:2101.07112 (2021).

\bibitem{Paesani17} S. Paesani, A. A. Gentile, R. Santagati, J. Wang, N.
Wiebe, D. P. Tew, J. L. O’Brien, and M. G. Thompson, Experimental Bayesian Quantum Phase Estimation on a Silicon Photonic Chip, Phys. Rev. Lett. 118, 100503 (2017).

\bibitem{Cimini19} V. Cimini, I. Gianani, N. Spagnolo, F. Leccese, F. Sciarrino, and M. Barbieri, Calibration of Quantum Sensors by Neural Networks, Phys. Rev. Lett. 123, 230502 (2019).

\bibitem{Nolan20} S. P. Nolan, A. Smerzi, and L. Pezz\'{e}, A machine learning approach to Bayesian parameter estimation, Eprint arXiv  arXiv:2006.02369 (2020).

\bibitem{Martina21}
S. Martina, S. Gherardini, and F. Caruso, Machine learning approach for quantum non-Markovian noise classification, Eprint arXiv:2101.03221 (2021).

\bibitem{Krenn16} M.Krenn, M. Malik, Mehul, R. Fickler, R. Lapkiewicz, Radek and A. Zeilinger, Automated Search for new Quantum Experiments, Phys. Rev. Lett. 116, 090405 (2016).

\bibitem{Melnikov18} A. A. Melnikov, H. P. Nautrup, M. Krenn, V. Dunjko, M. Tiersch, A. Zeilinger, and H. J. Briegel, Active learning machine learns to create new quantum experiments, Proc. National Acad. Sci. 115, 1221 (2018).

\bibitem{kadowaki1998quantum}
T. Kadowaki and H. Nishimori, Quantum annealing in the transverse Ising model, Physical Review E 58(5), 5355 (1998)

\bibitem{brooke2001tunable}
J. Brooke, T. F. Rosenbaum and G. Aeppli, Tunable quantum tunnelling of magnetic domain walls,  Nature 413.6856 (2001)

\bibitem{farhi2014quantum}
E. Farhi, J. Goldstone and S. Gutmann, A quantum approximate optimization algorithm, arXiv preprint arXiv:1411.4028 (2014)

\bibitem{peruzzo2014variational}
A. Peruzzo, J. McClean, P. Shadbolt, M. Yung, X. Zhou, P. J. Love, A. Aspuru-Guzik, and J. L. O'brien, A variational eigenvalue solver on a photonic quantum processor,Nature communications 5, 1 (2014)

\bibitem{kandala2017hardware}
A. Kandala, A. Mezzacapo, K. Temme, M. Takita, M. Brink, J. M. Chow and J. M. Gambetta, Hardware-efficient variational quantum eigensolver for small molecules and quantum magnets,Nature 549, 7671 (2017)

\bibitem{neven2008training}
H. Neven, V. S. Denchev, G. Rose and W. G. Macready, Training a binary classifier with the quantum adiabatic algorithm, arXiv preprint arXiv:0811.0416 (2008)

\bibitem{denchev2012robust}
V. S. Denchev, N. Ding, S. V. N. Vishwanathan and H. Neven, Robust classification with adiabatic quantum optimization, arXiv preprint arXiv:1205.1148 (2012).

\bibitem{pudenz2013quantum}
K.L Pudenz and D. A. Lidar, Quantum adiabatic machine learning, Quantum information processing 12, 5 (2013)

\bibitem{mott2017solving}
A. Mott, J. Job, J. Vlimant, D. Lidar and M. Spiropulu, Solving a Higgs optimization problem with quantum annealing for machine learning, Nature 550, 7676 (2017)

\bibitem{otterbach2017unsupervised}
J. S. Otterbach, R. Manenti, N. Alidoust, A. Bestwick, M. Block, B. Bloom, S. Caldwell et al., Unsupervised machine learning on a hybrid quantum computer, arXiv preprint arXiv:1712.05771 (2017).

\bibitem{vinci2020path}
W. Vinci, L. Buffoni, H. Sadeghi, A. Khoshaman, E. Andriyash, and M. H. Amin, A path towards quantum advantage in training deep generative models with quantum annealers, Machine Learning: Science and Technology 1, 4 (2020)

\bibitem{buffoni2021new}
L. Buffoni and F. Caruso, New trends in quantum machine learning, EPL 132, (2021)

\bibitem{preskill2018nisq}
J. Preskill, Quantum Computing in the NISQ era and beyond, Quantum 2, (2018).

\bibitem{suykens1999least}
J.A.K. Suykens and J. Vandewalle, Neural Processing Letters 9, 293 (1999).

\bibitem{pennylane}
V. Bergholm, J. Izaac, M. Schuld, C. Gogolin, C. Blank, K. McKiernan, and N. Killoran, Eprint arXiv:1811.04968.

\bibitem{rigettiqpu}
M.Reagor et al., Demonstration of universal parametric entangling gates on a multi-qubit lattice, Science advances 4.2, (2018).

\bibitem{smith2016practical}
R. S. Smith, M. J. Curtis and W. J. Zeng, A practical quantum instruction set architecture, arXiv:1608.03355 (2016).

\bibitem{atom_chip}
J. Petrovic, I. Herrera, P. Lombardi, F. Sch\"{a}fer, and F. S. Cataliotti, A multi-state interferometer on an atom chip, New J. Phys. 15, 043002 (2013).

\bibitem{opt_control}
C. Lovecchio, F. Sch\"{a}fer, S. Cherukattil, M. Al\`{i} Khan, I.Herrera, F. S. Cataliotti, T. Calarco, S. Montangero, and F.Caruso, Phys. Rev. A 93, 010304(R) (2016).

\bibitem{Stern_Gerlach}
W. Gerlach and O. Stern, Das magnetische moment dessilberatoms, Z. Phys. 9, 353 (1922).

\end{thebibliography}
\end{document}